\documentclass[preprint,aps,pra,showpacs,onecolumn]{revtex4}
\usepackage{color}

\usepackage{float}
\usepackage{amsmath}
\usepackage{graphicx}
\usepackage{amssymb}
\makeatletter

\usepackage{slashbox}
\RequirePackage{mathrsfs}
\usepackage{bm}
\usepackage{txfonts}

\newcommand{\uaa}{\raisebox{-0.3\height}{\includegraphics[height=1cm]{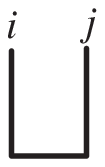}}}
\newcommand{\uab}{\raisebox{-0.3\height}{\includegraphics[height=1cm]{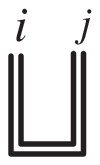}}}
\newcommand{\uba}{\raisebox{-0.3\height}{\includegraphics[height=1cm]{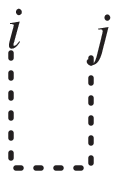}}}
\newcommand{\ubb}{\raisebox{-0.3\height}{\includegraphics[height=1cm]{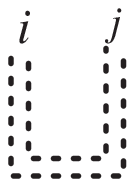}}}

\newcommand{\naa}{\raisebox{-0.3\height}{\includegraphics[height=1cm]{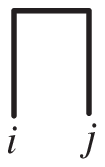}}}
\newcommand{\nab}{\raisebox{-0.3\height}{\includegraphics[height=1cm]{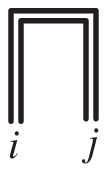}}}
\newcommand{\nba}{\raisebox{-0.3\height}{\includegraphics[height=1cm]{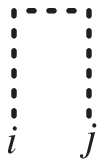}}}
\newcommand{\nbb}{\raisebox{-0.3\height}{\includegraphics[height=1cm]{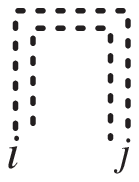}}}

\newcommand{\uaanaa}{\raisebox{-0.3\height}{\includegraphics[height=1.5cm]{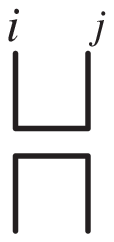}}}
\newcommand{\uabnab}{\raisebox{-0.3\height}{\includegraphics[height=1.5cm]{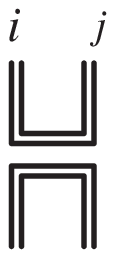}}}
\newcommand{\ubanba}{\raisebox{-0.3\height}{\includegraphics[height=1.5cm]{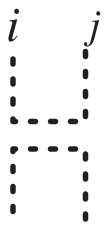}}}
\newcommand{\ubbnbb}{\raisebox{-0.3\height}{\includegraphics[height=1.5cm]{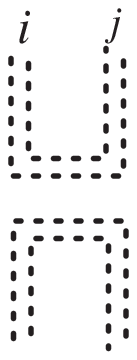}}}

\newcommand{\ha}{\raisebox{-0.3\height}{\includegraphics[height=1.5cm]{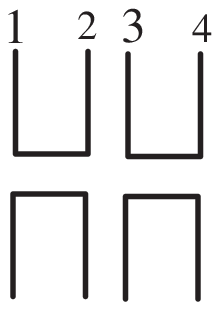}}}
\newcommand{\hb}{\raisebox{-0.3\height}{\includegraphics[height=1.5cm]{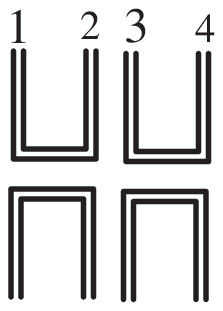}}}
\newcommand{\hc}{\raisebox{-0.3\height}{\includegraphics[height=1.5cm]{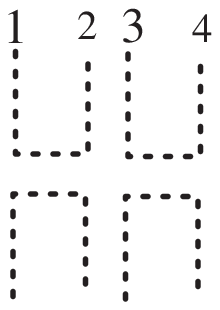}}}
\newcommand{\hd}{\raisebox{-0.3\height}{\includegraphics[height=1.5cm]{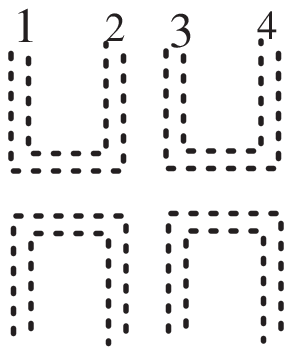}}}

\newcommand{\uoruaa}{\raisebox{-0.3\height}{\includegraphics[height=1cm]{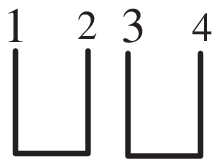}}}
\newcommand{\uoruab}{\raisebox{-0.3\height}{\includegraphics[height=1cm]{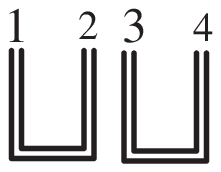}}}
\newcommand{\uoruba}{\raisebox{-0.3\height}{\includegraphics[height=1cm]{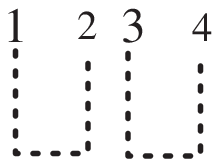}}}
\newcommand{\uorubb}{\raisebox{-0.3\height}{\includegraphics[height=1cm]{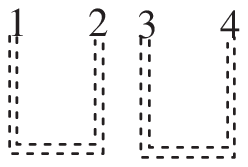}}}
\newcommand{\naaornaa}{\raisebox{-0.3\height}{\includegraphics[height=1cm]{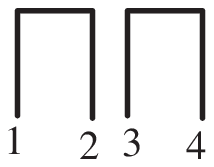}}}

\newcommand{\uaaanduaa}{\raisebox{-0.3\height}{\includegraphics[height=1cm]{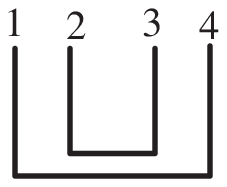}}}

\newcommand{\ubbanduab}{\raisebox{-0.3\height}{\includegraphics[height=1cm]{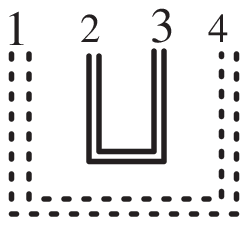}}}
\newcommand{\uaaanduaanaaornaa}{\raisebox{-0.3\height}{\includegraphics[height=1.6cm]{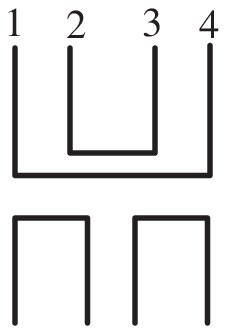}}}
\newcommand{\naaornaauaaanduaa}{\raisebox{-0.3\height}{\includegraphics[height=1cm]{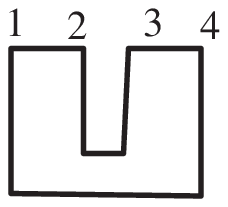}}}

\newcommand{\loopa}{\raisebox{-0.3\height}{\includegraphics[height=0.8cm]{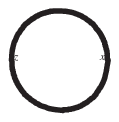}}}
\newcommand{\loopb}{\raisebox{-0.3\height}{\includegraphics[height=0.8cm]{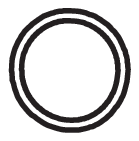}}}
\newcommand{\loopc}{\raisebox{-0.3\height}{\includegraphics[height=0.8cm]{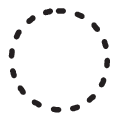}}}
\newcommand{\loopd}{\raisebox{-0.3\height}{\includegraphics[height=0.8cm]{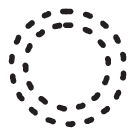}}}

\newcommand{\ket}[1]{|#1\rangle}
\newcommand{\bra}[1]{\langle#1|}
\newcommand{\inner}[2]{\langle#1|#2\rangle}

\makeatother

\begin{document}

%\date{}

\title{Quantum tunneling effect and quantum Zeno effect in a Topological system}

\author{Gangcheng Wang}
\author{Kang Xue\footnote{The Corresponding Author}}%
\email{youngcicada@163.com}
\author{Chunfang Sun}
\author{Chengcheng Zhou}
\author{Guijiao Du}
\affiliation{School of Physics, Northeast Normal University
\\Changchun 130024, People's Republic of China}

\date{\today}
\begin{abstract}
A spin interaction Hamiltonian for topological basis is constructed
in this paper. When we select proper parameters, this Hamiltonian
system can be simulated by a quantum double well potential system.
If the parameter $\Delta=0$, the topological system is equivalent to
two independent quantum wells. If the parameter $\Delta\neq 0$, the
system is equivalent to a double well potential with finite
potential barrier. The quantum tunneling effect and quantum Zeno
effect for this topological system are investigated in detail.
\end{abstract}


\maketitle

\section{Introduction}\label{sec1}
Topological quantum computing(TQC) is one of the most important
approaches to achieve a fault-tolerant quantum
computer\cite{qiybe1,zanardi,hqc,qiybe3}. The computation scheme
relies on the existence of topological states of matter whose
quasi-particle excitations are anyons, and they obey the braiding
statistics. Quantum information is stored in the states with
multiple quasi-particles with a topological degeneracy. Quantum gate
operations are implemented by braiding the quasi-particles. Thus a
quantum computer arises from the nonlocal encoding of the
multi-quasiparticle states, which makes them immune to errors caused
by local perturbations.

In Refs.\cite{kauffman1,zkg,zg} Kauffman \emph{et.al.} use the
CAP-CUP language reveal the relations between quantum mechanics and
topology. They said, ``the connection of quantum mechanics and
topology is an amplification of Dirac notation." For the following
convenience, we will introduce the Kauffman's CAP-CUP language. In
this paper, we refer to lines connecting lattice points in the top
row as $\ket{cup}$ states, those connecting lattice points in the
bottom row as $\bra{cap}$ states. We use $\bra{CAP}$ and $\ket{CUP}$
to denote the composition of caps and the composition of caps,
respectively. For example, the $\ket{CAP}$ state $\naaornaa$ is
composition of two $\bra{cap}$ states(\emph{i.e.} $\naa$), and the
$\bra{CUP}$ state $\uaaanduaa$ is composition of two $\ket{cup}$
states(\emph{i.e.} $\uaa$). By means of CAP-CUP states, we can
define operators and inner-product as following,
\begin{eqnarray*}
 && \hat{O} = \ket{\phi}\bra{\psi}=\left[\uaaanduaa\right]\cdot\left[\naaornaa\right]=\uaaanduaanaaornaa,\\
  &&
  Z(\psi,\phi)=\inner{\psi}{\phi}=\left[\naaornaa\right]\cdot\left[\uaaanduaa\right]=\naaornaauaaanduaa=\loopa.
\end{eqnarray*}
On the other hand, an experimental results for a small-scale
approximate evaluation of the Jones polynomial by nuclear magnetic
resonance(NMR) was presented in Ref.\cite{kauffman2}. The authors
could obtain the value of the Jones polynomial via measuring the
nuclear spin state of the molecule.

 Recently, in Ref.\cite{hu1} Ge \emph{et.al.} constructed a set of spin
realization topological basis. By means of Temperley-Lieb algebra
and Topological basis, the authors reduced the four-dimensional
Yang-Baxter Equation(YBE) into its two-dimensional form. Then they
point out that YBE can be tested in terms of quantum optics. On our
knowledge, there is not a Hamiltonian system describing the spin
realization of topological basis. Motivated by this, we will
construct a spin interaction Hamiltonian for this topological
system. Consequently, we can study physical properties of this
topological system, such as quantum tunneling effect(QTE) and
quantum Zeno effect(QZE).

This paper is organized as follows: in Sec.\ref{sec2}, we recall the
Temperley-Lieb algebra and topological basis. In Sec.\ref{sec3}, we
construct a topological Hamiltonian and obtain the eigen-system for
the topological system. In Sec.\ref{sec4}, we study the quantum
tunneling effect and quantum Zeno effect for this topological
system. We end with a summary.

\section{Temperley-Lieb algebra and Topological basis}\label{sec2}
 We first briefly review the theory of Temperley-Lieb(T-L) algebra\cite{TLA}.
Given a choice of topological parameter $d$ and a natural number
$m$, the T-L algebra $TL_{m}(d)$ is generated by
$\{I,U_{1},U_{2}\cdots U_{m-1}\}$ with the T-L algebra relations:
\begin{eqnarray}\label{tla}
\left\{
\begin{array}
[c]{lr} U_{i}^{2}=dU_{i} & 1\leq i \leq m-1,\\
U_{i}U_{i\pm1}U_{i}=U_{i} & 1\leq i\leq m, \\
U_{i}U_{j}=U_{j}U_{i} & \left\vert i-j\right\vert \geq2,
\end{array}
\right.
\end{eqnarray}
 where the notation $U_{i}\equiv U_{i,i+1}$ is used. The $U_{i}$
represents $1_{1}\otimes 1_{2}\otimes 1_{3}\cdots \otimes
1_{i-1}\otimes U\otimes 1_{i+2}\otimes \cdots \otimes 1_{m}$ , and
$1_{j}$ represents the unit matrix in the $j-$th space $\mathcal
{V}_{j}$. In addition, the T-L algebra is easily understood in terms
of knot diagrams, please refer to
Refs.\cite{kauffman1,kauffman2,kauffman3} and references therein.

The $4\times 4$ T-L matrix $U$ with $d=\sqrt{2}$ which satisfies T-L
algebra in Eqs.(\ref{tla}) takes the representation\cite{ckg,ckg2},
\begin{eqnarray}
  U &=& \frac{1}{\sqrt{2}}\left(
                            \begin{array}{cccc}
                              1 & 0 & 0 & iq^{-1} \\
                              0 & 1 & i\epsilon & 0 \\
                              0 & -i\epsilon & 1 & 0 \\
                              -iq & 0 & 0 & 1 \\
                            \end{array}
                          \right),
\end{eqnarray}
where $\epsilon=\pm$ and $q=e^{i\phi}$ with real $\phi$. For the
following convenience, we use the single solid lines, double solid
lines, single dash lines and double dash lines to distinguish
different topological states. Then we can introduce a set of
$\ket{cup}$ and $\bra{cap}$ states and their spin realization as,
\begin{eqnarray}\label{graph_state}
\begin{array}{l}
  \uaa=\sqrt{d}\ket{\psi_{d}^{(1)}}_{ij}=\sqrt{\frac{d}{2}}[\ket{\uparrow\uparrow}_{ij}+e^{-i\phi}\ket{\downarrow\downarrow}_{ij}]=\sqrt{d}[_{ij}\bra{\psi_{d}^{(1)}}]^{\dag}=\left[\naa \right]^{\dag},\\
  \uab=\sqrt{d}\ket{\psi_{d}^{(2)}}_{ij}=\sqrt{\frac{d}{2}}[\ket{\uparrow\downarrow}_{ij}-i\epsilon\ket{\downarrow\uparrow}_{ij}]=\sqrt{d}[_{ij}\bra{\psi_{d}^{(2)}}]^{\dag}=\left[\nab \right]^{\dag}, \\
  \uba=\sqrt{d}\ket{\psi_{0}^{(1)}}_{ij}=\sqrt{\frac{d}{2}}[\ket{\uparrow\uparrow}_{ij}-e^{-i\phi}\ket{\downarrow\downarrow}_{ij}]=\sqrt{d}[_{ij}\bra{\psi_{0}^{(1)}}]^{\dag}=\left[\nba\right]^{\dag}, \\
  \ubb=\sqrt{d}\ket{\psi_{0}^{(2)}}_{ij}=\sqrt{\frac{d}{2}}[\ket{\uparrow\downarrow}_{ij}+i\epsilon\ket{\downarrow\uparrow}_{ij}]=\sqrt{d}[_{ij}\bra{\psi_{0}^{(2)}}]^{\dag}=\left[\nbb
  \right]^{\dag},
\end{array}
\end{eqnarray}
where the notation $\uparrow$($\downarrow$) denotes
spin-up(spin-down), and the notation $\ket{\alpha\beta}_{ij}$ is the
abbreviated form of
$\ket{\alpha}_{i}\otimes\ket{\beta}_{j}$($\alpha,\beta=\uparrow,\downarrow$).
The topological parameter(the single loop)
$d=\loopa=\loopb=\loopc=\loopd=\sqrt{2}$ in this paper. In terms of
CAP-CUP language, the T-L matrix in Eq.(\ref{tla}) can be recast as
following,
\begin{equation}\label{graph_tla}
    U_{ij}=\uaanaa+\uabnab.
\end{equation}
Following Ge \emph{et.al.} in Ref.\cite{niu}, to reduce the $4\times
4$ Temperley-Lieb matrix, we can introduce a set of topological
basis states with four quasi-particles. The topological basis states
take the following form,
\begin{eqnarray}\label{graph_topologicalbasis}
  \begin{array}{l}
    \ket{e_{1}}=\frac{1}{d\sqrt{2}}\left[\uoruaa+\uoruab\right],\\
    \ket{e_{2}}=\frac{1}{d\sqrt{2}}\left[(1-i\epsilon e^{i\phi})\uaaanduaa+(i\epsilon+e^{-i\phi})\ubbanduab-\uoruaa-\uoruab\right],\\
    \ket{e_{3}}=\frac{1}{d\sqrt{2}}\left[\uoruaa-\uoruab\right],\\
    \ket{e_{4}}=\frac{1}{d\sqrt{2}}\left[(1-ie^{i\phi})\uaaanduaa-i\epsilon(1-ie^{-i\phi})\ubbanduab-\uoruaa+\uoruab\right].
  \end{array}
\end{eqnarray}
We can verify that this set of basis states are orthonormal
basis,(\emph{i.e.} $\inner{e_{i}}{e_{j}}=\delta_{ij}$). In fact,
topological bases $\ket{e_{2}}$ and $\ket{e_{4}}$ are equivalent to
the following simply form,
\begin{eqnarray}\label{graph_e2e4}
  \begin{array}{l}
    \ket{e_{2}}=\frac{-i\epsilon}{d\sqrt{2}}\left[e^{i\phi}\uoruba+e^{-i\phi}\uorubb\right]\\
    \ket{e_{4}}=\frac{i\epsilon}{d\sqrt{2}}\left[e^{i\phi}\uoruba-e^{-i\phi}\uorubb\right].
      \end{array}
\end{eqnarray}
 By means of Eqs.(\ref{graph_state}),
Eqs.(\ref{graph_tla}) and Eqs.(\ref{graph_topologicalbasis}), we can
verify that the T-L matrix can be reduced to two same $2D$
representations. The bases of subspace are
$\{\ket{e_{1}},\ket{e_{2}}\}$ and $\{\ket{e_{3}},\ket{e_{4}}\}$. The
$2D$ representations on the subspace $\{\ket{e_{1}},\ket{e_{2}}\}$
take the following form,
\begin{eqnarray}
  \begin{array}{cc}
    U_{A}=\left(
            \begin{array}{cc}
              d & 0 \\
              0 & 0 \\
            \end{array}
          \right);
     & U_{B}=\left(
               \begin{array}{cc}
                 d^{-1} & \sqrt{1-d^{-2}} \\
                 \sqrt{1-d^{-2}} & d-d^{-1} \\
               \end{array}
             \right),
  \end{array}
\end{eqnarray}
where $(U_{A})_{ij}=\bra{e_{i}}U_{12}\ket{e_{j}}$ and
$(U_{B})_{ij}=\bra{e_{i}}U_{23}\ket{e_{j}}$($i,j=1,2$). We can
verify $U_{A}$ and $U_{B}$ satisfy the $2D$ T-L relations,
$U_{A}^{2}=dU_{A}$, $U_{B}^{2}=dU_{B}$, $U_{A}U_{B}U_{A}=U_{A}$ and
$U_{B}U_{A}U_{B}=U_{B}$. This $2D$ form T-L matrices can be
constructed by using the conformal field theory, and have been
applied to the fractional quantum Hall effect. Please see
Refs.\cite{niu,wan} and references therein.

\section{Hamiltonian and Eigen-system for the topological
system}\label{sec3} For investigating the physical effect of the
spin realization of the topological basis, we should construct a
Hamiltonian for the topological system. The Hamiltonian for our
system reads,
\begin{eqnarray}\label{graph_hamiltonian}
  \hat{H} &=& Jd^{-2}\left[(1+\Delta)\left(\ha+4\hc\right)+(1-\Delta)\left(\hb+4\hd\right)\right] \nonumber\\
   &=& Jd^{-2}[(1+\Delta)(U_{12}^{(1)}U_{34}^{(1)}+4\tilde{U}_{12}^{(1)}\tilde{U}_{34}^{(1)})+(1-\Delta)(U_{12}^{(2)}U_{34}^{(2)}+4\tilde{U}_{12}^{(2)}\tilde{U}_{34}^{(2)})]
\end{eqnarray}
where $U_{ij}^{(1)}=\uaanaa$, $U_{ij}^{(2)}=\ubanba$,
$\tilde{U}_{ij}^{(1)}=\uabnab$ and $\tilde{U}_{ij}^{(2)}=\ubbnbb$.
If $\Delta=0$, we can verify that $\ket{e_{i}}(i=1,2,3,4)$ are
eigenstates of the Hamiltonian. The corresponding eigen-energy  are
$E_{1}=E_{3}=J$ and $E_{2}=E_{4}=4J$. The energy levels of the
Hamiltonian $\hat{H}$ in Eq.(\ref{graph_hamiltonian}) when
$\Delta=0$ are doubly degenerate. If the parameter $\Delta$ is a
finite real number, then the degenerate energy levels $J$ and $4J$
will split into two non-degenerate energy levels. Then the
eigen-states for the Hamiltonian $\hat{H}$ in
Eq.(\ref{graph_hamiltonian}) are found to be,
\begin{eqnarray}
\begin{array}{c}
  \ket{E_{1}^{\pm}}=\frac{1}{\sqrt{2}}(\ket{e_{1}}\pm\ket{e_{3}}) \\
  \ket{E_{2}^{\pm}}=\frac{1}{\sqrt{2}}(\ket{e_{2}}\pm\ket{e_{4}}),
\end{array}
\end{eqnarray}
with the corresponding eigen-values $E_{1}^{\pm}=J(1\pm\Delta)$ and
$E_{2}^{\pm}=4J(1\pm\Delta)$. If we set
$J=\frac{\hbar^{2}\pi^{2}}{2m(L-a)^{2}}$ and $\Delta=\frac{4e^{-2\xi
a}}{\xi(L-a)}$ with $\xi=\frac{\sqrt{2mV_{0}}}{\hbar}$, then the
Hamiltonian $H$ in Eq.(\ref{graph_hamiltonian}) is equivalent to a
double well system with the following potential function,
\begin{eqnarray}
  V(x) &=& \left\{\begin{array}{lr}
                   V_{0} & |x|\leq a \\
                   0 & a<|x|<L \\
                   +\infty & |x|>L.
                 \end{array}\right.
\end{eqnarray}
If potential barrier is infinity(i.e. $V_{0}\rightarrow +\infty$),
then $\Delta\rightarrow 0$. In this case, the system is equivalent
to two independent potential wells. The topological bases
$\ket{e_{1}}$ and $\ket{e_{3}}$ are equivalent to the ground states
for the left and right potential well, correspondingly. The
topological bases $\ket{e_{2}}$ and $\ket{e_{4}}$ are equivalent to
the first excited states for the left and right potential well,
correspondingly.

\section{Quantum tunneling effect and Quantum Zeno effect}\label{sec4}

In this section, we will consider two physical effects, Quantum
tunneling effect\cite{landau,razavy} and quantum Zeno
effect\cite{misra} in the topological system.

\subsection{Quantum tunneling effect}\label{sec4:sub1}
For our convenience, we will focus on the subspace
$\{\ket{e_{1}},\ket{e_{3}}\}$. Quantum mechanics predict that even
if the system has an energy less than the barrier height, it has a
finite probability to tunnel the barrier to the other side of the
potential barrier. By means of the topological system, we can study
the process of quantum dynamical tunneling. At $t=0$, we suppose
that the initial state is $\ket{e_{1}}$, and $\ket{e_{1}}$ can be
recast as a superposition of the eigenstates of $H$ in
Eq.(\ref{graph_hamiltonian}),
$\ket{e_{1}}=\frac{1}{\sqrt{2}}(\ket{E_{1}^{+}}+\ket{E_{1}^{-}})$.
Then the evolution of the system at any time $t$ is given by,
\begin{eqnarray}\label{psi_t}
  \ket{\psi(t)}&=&\frac{1}{\sqrt{2}}e^{-i\omega_{+} t}(\ket{E_{1}^{+}}+e^{i\delta
  t}\ket{E_{1}^{-}}),
\end{eqnarray}
where $\omega_{+}=J(1+\Delta)/\hbar$, $\omega_{-}=J(1-\Delta)/\hbar$
and $\delta=\omega_{+}-\omega_{-}$. So that,
\begin{eqnarray}\label{p_e1e3}
 \begin{array}{cc}
   P(\ket{e_{1}})=\cos^{2}(\delta t/2), & P(\ket{e_{3}})=\sin^{2}(\delta t/2),
 \end{array}
\end{eqnarray}
with $P(\ket{e_{1}})$ and $P(\ket{e_{3}})$ being probability for the
topological bases $\ket{e_{1}}$ and $\ket{e_{3}}$, correspondingly.
So, the $\ket{\psi(t)}$ will oscillate between topological bases
$\ket{e_{1}}$ and $\ket{e_{3}}$. According to the
Eqs.(\ref{p_e1e3}), then we can obtain the tunneling time is
$\tau=\pi/\delta$.

\subsection{Quantum Zeno effect}\label{sec4:sub2}

Quantum measurement will alter the dynamics of the system. In 1977,
Misra et.al. show that an unstable particle will never be found to
decay if it is continuously observed. They called it quantum Zeno
effect(QZE). Now, we show that in our topological system, the QZE
can be observed. In Sec.\ref{sec4:sub1}, we obtain the tunneling
time is $\tau=\pi/\delta$. Assume the time $\tau$ is divided into
$n$ equal parts. After time $\tau/n$, the probability of founding
the topological basis $\ket{e_{1}}$ is,
\begin{equation}
    P_{\ket{e_{1}}}(\tau/n)=\cos^{2}\frac{\pi}{2n}.
\end{equation}
After $n$ times measurements, the probability reads,
\begin{equation}
    P_{\ket{e_{1}}}^{n}=(\cos^{2}\frac{\pi}{2n})^{n}.
\end{equation}
As $n$ is large enough(i.e. $n\gg 1$), the probability approximately
is,
\begin{equation}
    P_{\ket{e_{1}}}^{n}\approx e^{-\frac{\pi^{2}}{4n}}\rightarrow 1.
\end{equation}
Then this is just the QZE.

\section{Summary}

In summary, we have presented the Hamiltonian to the topological
system, and we obtain the eigen-system for this system. When we
select proper parameters, this topological system can be simulated
by a quantum double well potential system. Based on which, quantum
tunneling effect and quantum Zeno effect have been studied in
detail. Let us make a summary to end this paper. Firstly, quantum
tunneling effect can be observed in this topological system, and the
tunneling time is $\tau=\pi/\delta$. Secondly, if we divide the
tunneling time $\tau$ into $n$ equal parts(\emph{i.e.}, each time
interval is $\tau/n$), after $n$ times measurements, the quantum
Zeno effect can occur in this topological system.

 Eventually, people have
currently found that topological basis has some important physical
applications in topological quantum computation, quantum
entanglement and topological quantum teleportation, how to reveal
the role of the topological parameter $d$ is an interesting and
significant topic. We shall investigate this subject subsequently.

\section*{Acknowledgments}
This work was supported by NSF of China (Grants No. 10875026) and
the Fundamental Research Funds for the Central Universities(Grants
No. 09SSXT026)

\end{document}